# A Comparative Study of Joint-SNVs Analysis Methods and Detection of Susceptibility Genes for Gastric Cancer in Korean Population

Jin-Xiong Lv[1], Shikui Tu[1], Lei Xu[1, 2*],

[1]Center for Cognitive Machines and Computational Health, and Department of Computer Science and Engineering, Shanghai Jiao Tong University,
200240 Shanghai, China

[2]Department of Computer Science and Engineering, The Chinese University of Hong Kong,
Hong Kong SAR, China
{lvjinxiong, tushikui, leixu}@sjtu.edu.cn

**Abstract.** Many joint-SNVs (single-nucleotide variants) analysis methods were proposed to tackle the 'missing heritability' problem, which emphasizes that the joint genetic variants can explain more heritability of traits and diseases. However, there is still lack of a systematic comparison and investigation on the relative strengths and weaknesses of these methods. In this paper, we evaluated their performance on extensive simulated data generated by varying sample size, linkage disequilibrium (LD), odds ratios (OR), and minor allele frequency (MAF), which aims to cover almost all scenarios encountered in practical applications. Results indicated that a method called Statistics-space Boundary Based Test (S-space BBT) showed stronger detection power than other methods. Results on a real dataset of gastric cancer for Korean population also validate the effectiveness of the S-space BBT method.

## 1 Introduction

The GWAS has made tremendous success based on the hypothesis 'Common Disease, Common Variant (CDCV)' [1], yet common variants identified via the GWAS only explained a small fraction of the heritability factors owing to two aspects. First, the traditional GWAS only focuses on the common variants to the common diseases, while the rare variants also make contributions to the common diseases in the light of 'Common Disease, Rare Variant (CDRV)' [2], and it is defined through the MAF (1% ≤ MAF ≤ 5%); second, it aims to detect the single genetic variants to the diseases while neglects the combined effect of SNVs [3]. The 'next generation' sequencing technologies facilitate the detection for rare variants contributing to the complex

---

diseases. However, interesting rare variants have difficulty in being captured owing to the insufficient sample size.

In view of this, investigators have proposed many joint-SNVs analysis methods to solve them. These methods can be divided into three categories via the way to obtaining the corresponding statistics. The first road is the 'projection'. We transform the statistics vector into one statistic for simplified calculation of *P*-value. Thus, it is crucial to define suitable 'projection' matrix. For instance, the Hotelling's T square test transforms the difference of mean vectors for two populations into the Hotelling's T square statistic by multiplying the inverse of covariance matrix. However, accurate estimation of the covariance matrix depends on the large sample size and the low missing rate. On the basis of CDRV, some methods collapse or sum up all SNVs in a unit into a single one to discover the accumulation effect of rare variants. Here, the 'projection' matrix is diagonal. These methods can be divided into two groups according to whether the 'projection' matrix is the identity matrix or not. The two groups are named after burden test and non-burden test. The burden test assumes that SNVs contribute to the unit equally, while the non-burden test does not. The second is the 'combination' in the probability space [4]. The Fisher's method combined the *P*-value of each hypothesis into the Chi-square statistic in linear form, but much information is lost. And the third road is that we perform the multivariate test in the high dimensional space directly, so that it can break through the two limitations that the existing methods suffered from [5]. First, these existing methods regardless of the relationship between the dimensions; Second, the direction for each component is not taken into consideration. It is of note that S-space BBT is the representative method in the third road, and the comparative study involved it is still absent both in the large-scale simulation dataset and real-world dataset.

In this paper, we first introduced six representative methods, and then performed the comparative study in considering the varying sample size, OR, LD and MAF. The simulation results showed that (1) all the involved methods obtain stronger detection power with the sample size increasing. S-space BBT and SKATO are more sensitive than other four methods; (2) S-space BBT has stronger detection power than other methods under different OR, LD and MAF; (3) S-space BBT almost obtains smaller *P*-value compared with other methods in the real-world datasets. All above indicate that the S-space BBT plays an important role for joint-SNVs analysis. Thus, we applied it to a dataset of gastric cancer for Korean population and obtained a susceptibility gene list. The literature survey for selected genes was conducted to validate the effectiveness of S-space BBT. As a result, we provided the biomarker list and anticipated that it can be the reference for the gastric cancer study.

## 2 Representative Methods

### 2.1 Hotelling's T Square Test

The Hotelling's T square distribution is the generalization of the Student's t-distribution. Given two populations and they follow the independent multivariate

normal distributions with same mean and covariance. The Hotelling's T square statistic is defined as:

$$t^2 = \frac{mn}{m+n}(\tilde{\mathbf{x}} - \tilde{\mathbf{y}})' \hat{\Sigma}^{-1} (\tilde{\mathbf{x}} - \tilde{\mathbf{y}}) \sim T^2(p, m+n-2)$$

$$\frac{m+n-p-1}{(m+n-2)p} t^2 \sim F(p, m+n-1-p)$$

(1)

where the $m$ and $n$ are the size of two populations, $p$ is the number of variates, $\tilde{\mathbf{x}}$ and $\tilde{\mathbf{y}}$ are the sample means and the $\hat{\Sigma}$ indicates the covariance matrix. In order to calculate the *P*-value, we often transform it into F statistics.

As for the case-control study, the Hotelling's T square test obtains more accurate *P*-value when the sample size is large and the missing rate for genotype data is low, because both lead to the precise estimation for the covariance matrix.

## 2.2 Sumstat Test

The sumstat test, one kind of the burden test, treats the SNVs equally in the unit and adds all of the statistics from each SNV together to conduct the hypothesis test. It can enhance the power in considering the existence of rare variants. But it ignores the effect direction and the magnitude effect of SNVs. When the SNVs have same effect direction and the magnitude, the sumstat test obtained the better performance. The effect direction is defined via the OR, when the OR > 1, the direction is deleterious, otherwise, the direction is protective.

## 2.3 The Sequence Kernel Association Test (SKAT) and its Optimal Version

For the regression model, we test whether the unit has influence on phenotype under the null hypothesis as described in the Eq.(1).

$$H_0: \boldsymbol{\beta} = 0 \tag{2}$$

where the $\boldsymbol{\beta}$ indicates the coefficients vector and the null hypothesis means that the corresponding SNV is not associated with the phenotype. The SKAT assumes each $\beta_j$ follows an arbitrary distribution with a mean of zero and a variance of $w_j\tau$ and then tests the null hypothesis $H_0: \tau = 0$ where the $w_j$ is the prespecified weight. It obtains the variance-component score statistics which take the direction of $\beta$ into consideration.

The optimal version of SKAT (SKATO) [6] combined burden statistic $Q_{burden}$ and SKAT statistic $Q_{SKAT}$ into the SKATO statistic $Q_\rho$ in linear form. SKATO statistic is described as followed:

$$Q_\rho = (1-\rho)Q_{SKAT} + Q_{burden} \tag{3}$$

where $\rho$ indicates pair-wise correlation among $\beta_j$ in Eq.(1).

Both of them belong to the non-burden test, not only taking the effect direction into account but also the magnitude of effect. So compared with the burden test, they are more robust.

### 2.4 Fisher's Combined Test

The Fisher's combined test combines the *P*-value from each test into the Chi-square statistic assuming the hypotheses are independent. The formula is defined as:

$$\chi^2_{2k} \sim -2\sum_{i=1}^{k} \ln(p_i) \tag{4}$$

where the *k* is the number of the hypotheses and the $p_i$ is the *P*-value obtained from the *i*-th hypothesis.

It suffers from the poor performance in joint-SNVs analysis owing to the information loss (e.g., LD, effect direction and so on). If there are many causal variants in the unit, the Fisher's combined method can achieve better performance.

### 2.5 Statistic-space Boundary Based Test

The above tests reject the H$_0$ as long as at least one of dimensions is rejected, and they ignore the roles of dimensions and their combination just as described in the Fig 3(a) of [5]. The S-space BBT is one of the directional test and is described in the Fig 6(a) of [5]. The way to achieving the combination is also given in Eq.(13)-(19) of [5].

The implementation of S-space BBT has been described in details in the Tab. 6 of [4]. Here, we give some key points of it. First, we directly use the boundary to form the rejection domain in the statistic space as followed:

$$\Gamma(\tilde{\mathbf{s}}) = \left\{ \mathbf{s} : (\mathbf{s} - \tilde{\mathbf{s}})' diag(\mathbf{sign}(\tilde{\mathbf{s}})) > \mathbf{0} \right\} \tag{5}$$

where $\mathbf{sign}(\mathbf{s}) = [sign(s_1), \cdots, sign(s_m)]'$ with $sign(v) = \frac{v}{|v|}$. Second, the *P*-value is calculated by the permutation test (see (65) in [4]). Third, the principle component analysis is performed to remove the second-order dependence. Forth, we adopt the posteriori version of the *P*-value for reduction of the background disturbance (see (93) of in [4]).

We have analyzed the application scenarios for the six methods in theory. The related computation have shown three points in the [7]. First, the six methods except for the S-space BBT swamp the significant SNVs. Second, burden test is powerful under the same effect direction. While the SKAT/SKATO is suitable for the different effect direction. Third, S-space BBT has stronger detection power in different MAF, LD and OR. In this paper, we adopted the statistic power to evaluate the six methods under the sample size, LD, OR and MAF in the simulation experiments. The detection power is defined as the proportion of true positive results. They were also evaluated on the real-world datasets.

## 3 Simulation Experiments

### 3.1 Simulation Framework

In order to compare the power of different approaches under various conditions, we use the simulation tool of PLINK software [8] to generate large simulation datasets. The number of SNVs in the joint unit is 10, which is composed of 5 causal variants and corresponding 5 observed markers. As a result, we obtained the simulation data of 10 SNVs on 100 cases vs. 100 controls, 500 cases vs. 500 controls and 1000 cases vs. 1000 controls in a stochastic way. Other parameter settings for the simulation datasets were described in the Table 1. Note that the LD in the Table 1 is calculated between causal variant and corresponding observed marker, so we call it the incomplete LD. Besides, we produced 1000 replicates for each dataset for power computation and set the threshold $\alpha = 0.05$. In the [9], the detection power was estimated as the proportion of $P$-value $\leq \alpha$ among the 1000 replicates.

Hotelling's T square test, Fisher's combined test and S-space BBT were implemented by the MATLAB. We adopted the SKATBinary function in the SKAT package of the R software to perform SKAT and SKATO. Sumstat test was performed by the PLINK/seq software with 100000 times of permutation.

**Table 1.** Parameter settings for simulation datasets

| Conditions | DatasetID | $OR_{het}$ | $OR_{hom}$ | MAF | Marker / causal variant LD |
|---|---|---|---|---|---|
| LD | Dataset1 | 1.2 | 2.4 | 0.05 | 0.4 |
|  | Dataset2 | 1.2 | 2.4 | 0.05 | 0.96 |
| OR | Dataset3 | 1.1 | 2.2 | 0.05 | 0.8 |
|  | Dataset4 | 1.2 | 2.4 | 0.05 | 0.8 |
|  | Dataset5 | 1.3 | 2.6 | 0.05 | 0.8 |
| MAF | Dataset6 | 1.2 | 2.4 | 0.01 | 0.8 |
|  | Dataset7 | 1.2 | 2.4 | 0.03 | 0.8 |

Note: $OR_{het}$ indicates the odds ratio for heterozygote causal variants.
$OR_{hom}$ indicates the odds ratio for homozygote causal variants.

### 3.2 Simulation Results

#### 3.2.1 Linkage Disequilibrium

We first focus on the effect of the incomplete LD on each method. The linkage disequilibrium is the correlation between two SNVs and can be measured with the correlation coefficient [10]. The results were shown in the Fig. 1.

All of the methods achieved higher detection power with the sample size increasing. In particular, the accurate estimation of the covariance matrix may account for the improvement for Hotelling's T square test, SKAT and SKATO. The S-space BBT obtained the best performance among the six methods. The SKATO obtained stronger detection power than sumstat test and SKAT. In conclusion, the power of the six methods is almost constant in different incomplete LD.

### 3.2.2 Odds Ratio

The odds ratio is utilized to quantify the relationship between property A and property B in a given population. In GWAS, it quantifies the impact that one allele has on disease. When the OR > 1, the SNV is defined as deleterious one, which means that the more frequent the allele of SNV appears, the more likely to get sick. Conversely, when the OR < 1, the SNV is defined as protective one.

As shown in the Fig. 2, the power for each method enhances as the OR increasing, and SKATO is more sensitive than other methods. The sensitivity indicates the growth rate of detection power under different conditions. When the OR is large enough and the sample size is 2000, all methods achieved at least 85% power. The S-space BBT still keep the highest power in the different OR.

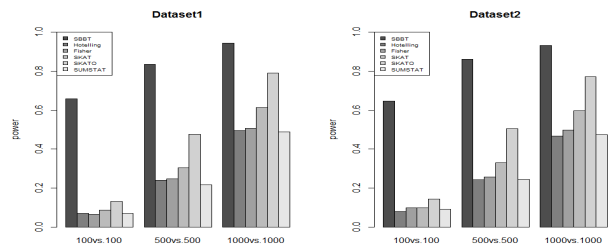

**Fig. 1.** Power comparison under different incomplete LD

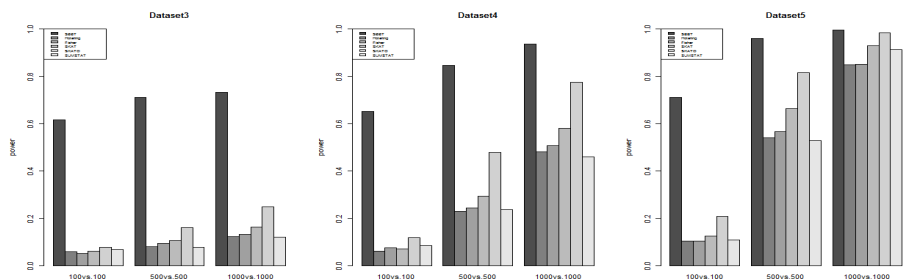

**Fig. 2.** Power comparison under different OR

### 3.2.3 Minor Allele Frequency

One site has two alleles (e.g. 'A' and 'a') in general. The frequency of second most common allele is the minor allele frequency in a given population. The rare variants are defined by the minor allele frequency. Based on the CDRV, the rare variants play a crucial role in genetic susceptibility to common diseases [2].

As described in the Fig. 3, all the six methods obtained stronger power with the MAF increasing, and each achieved greater sensitivity. It is of note that the S-space BBT kept the better performance ($power_{average} \geq 60\%$) when the MAF = 1%, while other methods achieved less than 20% average power. The Fisher's combined method achieved 0.2% power when the sample size is 100 vs. 100 and the MAF = 1%.

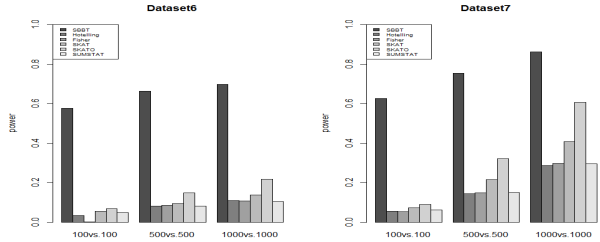

**Fig. 3.** Power comparison under different MAF

## 4   Gastric Cancer Study

### 4.1   Quality Control

The gastric cancer is the fifth most common malignancy in the world, especially in the Korea [11]. The selected dataset is associated with the gastric cancer for Korean population from the GEO database (Gene Expression Omnibus, ID: GSE58356). 319283 probes make up the dataset with the sample size of 683 controls and 329 cases.

As for the quality control, we took the Hardy-Weinberg's equilibrium and the missing rate into consideration. The Hardy-Weinberg law states that the allele and genotype frequency in a population will remain constant generation after generation. It is essential to regard the Hardy-Weinberg's equilibrium as one of measures in the quality control owing to the identification of questionable genotypes [12]. The threshold of the missing rate is set to 5%, and the threshold of the Hardy-Weinberg's equilibrium is set to 1.00E-04. After quality control and removing the duplicate probes, 54988 SNVs were remained. We regarded the gene as a unit and obtained 14709 units to conduct the joint-SNVs analysis via the S-space BBT.

### 4.2 Evaluation on the Real-world Dataset

To overcome some limitations (e.g., the LD, existence of the causal variants and so on) of the simulation experiments, we made efforts to search for the SNVs that are not only generally recognized but also can be found in the published SNVs datasets. Finally, 3 significant SNVs were found. Then all the six methods were performed for them and the results were shown in the Table 2.

**Table 2.** Results of three benchmarks for the six methods

| Gene  | SBBT     | Hot      | Fis      | SKAT    | SKATO   | SUM     |
|-------|----------|----------|----------|---------|---------|---------|
| PSCA  | 8.90E-10 | 1.23E-06 | 1.65E-12 | 4.49E-01| 4.49E-01| 1.30E-03|
| ANK3  | 5.62E-08 | 4.83E-04 | 5.66E-03 | 7.43E-02| 1.09E-01| 6.85E-03|
| PALB2 | 7.96E-07 | 2.09E-03 | 1.13E-03 | 3.04E-01| 3.28E-01| 5.35E-03|

Note: Hot indicates the Hotelling's T square tests; Fis means the Fisher's combined test; SUM indicates the sumstat test.

The well-known rs2294008 in *PSCA* (prostate stem cell antigen) [13,14] is involved in the GSE58356 dataset, so that it can be the benchmark for comparative study. It was also identified in our gastric cancer study (*P*-value = 1.12E-07, OR = 1.66). In our dataset, the *PSCA* contains three SNVs, so the Hotelling's T square test obtained better performance. There is other SNV (rs1045531) whose *P*-value is 7.73E-08 in the *PSCA*, which leads to the small *P*-value for Fisher's combined method. S-space BBT and Fisher's combined method maintained the significance of the causal variant (rs2294008) while others did not.

The rs1938526 and rs420259 are found in the GSE71443 dataset consisting of 65 bipolar disorder patients and 74 controls. The dataset contains no missing value. We adopted similar quality control as the GSE58356 did.

The rs1938526 of the *ANK3* (Ankyrin 3) is the susceptibility locus for the bipolar disorder in [15]. For the *ANK3*, there are 235 SNVs involved, which may result in the inaccuracy computation of joint *P*-values owing to the small size of population. Thus, we selected SNVs located in the upstream and downstream 20kb of rs1938526 to make up the computational unit, and 15 SNVs were remained. The smallest single locus *P*-value is 1.21E-05, while the *P*-value of rs1938526 ranks second (*P*-value = 0.06, OR = 1.58). The Hotelling's T square test obtained the smaller *P*-value compared to the other methods except for the S-space BBT owing to no missing value.

The rs420259 in *PALB2* (Partner And Localizer Of *BRCA2*) is also regarded as the meaningful SNV for the bipolar disorder [16]. For the *PALB2*, three SNVs are involved in. The rs420259 has the smallest *P*-value (*P*-value = 2.95E-03, OR = 0.42). Fisher's method, Hotelling's T square test and sumstat test achieved similar *P*-value.

In conclusion, the S-space BBT achieved smaller *P*-value compared with other methods for the *ANK3* and *PALB2*. As for the *PSCA*, the *P*-value of S-space BBT is also significant ($p_{SBBT} \leq 2.50\text{E-}06$). The SKAT and the SKATO performed worst for the three genes, which might result from the small sample size.

### 4.3 Literature Survey for Top 20 Associated Genes of Gastric Cancer

It is of note that we mainly focus on the combined effect of SNVs in the comparative study. Thus, some genes would be neglected owing to two points. First, smallest *P*-values of SNVs in these genes are smaller than 5.00E-08, which indicates that the traditional GWAS can detect them; second, there is only one SNV in the gene. Then, we selected top 20 genes detected via S-space BBT to conduct the literature survey. The search result was shown in the **Appendix** in detail. Further, we divided the genes into three groups. C group means those genes related to gastric cancer, B group indicates those related to other kinds of cancers and A group is other cases. In summary, there are 65% genes associated with the gastric cancer and other kinds of cancers. It indicates that the S-space BBT is reliable in the joint-SNVs analysis.

## 5 Conclusion

We conducted a comparative study on the main threads of joint-SNVs analysis methods in considering the sample size, LD, OR and MAF. The simulation experiments were designed to show that the S-space BBT has stronger detection power compared with other involved methods in different conditions. The simulation results showed that the S-space BBT plays an crucial role in detection of the susceptibility genes. More generally, we evaluated them on the real-world dataset and reached the same conclusion. Thus, we applied the S-space BBT to the dataset of gastric cancer for Korean population and obtained 20 significant genes. In order to validate the efficiency of the S-space BBT, we conducted literature survey for the top 20 genes, of which 65% are associated with the gastric cancer and other kinds of cancers. The prevalence of many diseases is low, which leads to the sample disequilibrium problem in statistical tests. The reactions of different joint-SNVs analysis methods to the problem might be an interesting issue, further, it is essential for investigators to propose novel methods to solve it.

## Acknowledgements

This work was supported by the Zhi-Yuan chair professorship start-up grant (WF220103010) from Shanghai Jiao Tong University.